\documentclass[onecolumn,superscriptaddress,twoside,pra,nofootinbib]{revtex4}
\usepackage{graphicx}
\usepackage[latin1]{inputenc}
\usepackage[english]{babel} 
\usepackage{thumbpdf}
\usepackage{subfigure}    
\usepackage{color}
\usepackage{amsmath}
\usepackage{amsfonts}
\usepackage{amssymb}
\usepackage{amsbsy}
\usepackage{bbm}
\usepackage{bm}
\usepackage{epsfig}
\usepackage{epstopdf}
\usepackage{dsfont}
\usepackage{color}
\usepackage[colorlinks]{hyperref}
\usepackage[figure,table]{hypcap}
\usepackage{enumerate}
\hypersetup{
	bookmarksnumbered,
	pdfstartview={FitH},
	citecolor={darkgreen},
	linkcolor={darkblue},
	urlcolor={darkblue},
	pdfpagemode={UseOutlines}}
\definecolor{darkgreen}{RGB}{50,190,50}
\definecolor{darkblue}{RGB}{0,0,190}
\definecolor{darkred}{RGB}{238,0,0}
\usepackage{soul}

\newcommand{\bra}[1]{\langle #1|}
\newcommand{\ket}[1]{|#1\rangle}

\newcommand{\apicisx}{\textquotedblleft}
\newcommand{\apicidx}{\textquotedblright}
\newcommand{\G}{\mathcal{G}}

\newcommand{\id}{\mathbbm{1}}

\newcommand{\gr}[1]{\boldsymbol{#1}}
\newcommand{\be}{\begin{equation}}
\newcommand{\ee}{\end{equation}}
\newcommand{\bea}{\begin{eqnarray}}
\newcommand{\eea}{\end{eqnarray}}
\newcommand{\ketbra}[2]{\vert #1 \rangle \! \langle #2 \vert}

\newtheorem{defn}{Definition}
\newtheorem{prop}{Proposition}

\begin{document}

\title{Quantum State Transfer through Noisy Quantum Cellular Automata}

\author{Michele Avalle}
\email{michele.avalle.11@ucl.ac.uk}
\author{Marco G. Genoni}
\author{Alessio Serafini}
\affiliation{Department of physics and Astronomy, University College London, Gower Street, London, WC1E 6BT}
\date{\today}

\begin{abstract}
We model the transport of an unknown quantum state on one dimensional qubit lattices by means of a quantum
cellular automata evolution.
We do this by first introducing a class of discrete noisy dynamics, in the first excitation sector,
in which a wide group of classical stochastic dynamics is embedded within the more general formalism of quantum operations.
We then extend the Hilbert space of the system to accommodate a global vacuum state, thus allowing for the transport
of initial on-site coherences besides excitations, 
and determine the dynamical constraints that define the class of noisy quantum cellular automata in this subspace.
We then study the transport performance through numerical simulations, showing that for some instances of the dynamics
perfect quantum state transfer is attainable.
Our approach provides one with a natural description of both unitary and open quantum evolutions,
where the homogeneity and locality of interactions 
allow one to take into account several forms of quantum noise in a plausible scenario.    
\end{abstract}

%keywords{
%
%}

%\pacs{
%
%}

\maketitle

%%%%%%%%%%%%%%%%%%%%%%%%%%%%%%%%%%%%%%%%%%%%%%%%%%%%%%%%%%%%%%%%%%%%%%%%%%% INTRODUCTION
\section{Introduction}
The transfer of a quantum state between two or more different points in space is an 
important task in quantum information processing. 
To this end, arrays of permanently coupled quantum systems in which quantum state transfer
is accomplished through free dynamical evolution  - also knows as \textit{spin chains} \cite{Bose1,Bose2} -
have drawn substantial attention in the last ten years, up to the point of establishing 
themselves as a self standing area of study \cite{cristal1,cristal2,Albanese,Vittorio,Daniel1,Daniel2},

A class of systems intimately related to spin chains \cite{cristal1,cristal2} is represented by quantum walks,
the (either continuous or discrete time evolving) quantum counterpart to random walks on a lattice \cite{Aharonov1,Aharonov2,Kempe}.
Although originally introduced to investigate quantum speedups over classical algorithms \cite{Childs,Farhi},
quantum walks have been applied to the problem of state transport as well, proving the capability to achieve
perfect state transfer, even though usually under the engineering of rather restrictive dynamical 
protocols (e.g. \cite{cristal1,cristal2,turchi}; see also \cite{Kendon} and references therein).

Here we study coherent quantum state transfer on lattices governed by
quantum cellular automata \cite{Schumacher,ArrighiQCA}.
These systems are substancially different but nevertheless in a sense closely related
to spin chains and quantum walks \cite{Schumacher}. 
Quantum cellular automata (QCA) can be described as a set of quantum
systems (cells) on a d-dimensional regular lattice evolving in discrete 
time steps according to a certain fixed transition rule.
This rule has to be local (information cannot travel faster
than some fixed number of cells per time step) and the global dynamics must be
translationally invariant, such that the physics is homogeneous across the lattice.
Because of this \apicisx physics-like\apicidx structure
and the inherent parallelism, QCA were first envisaged as potentially versatile and interesting models
for quantum computation and quantum simulation \cite{Margolus,Feynmann}.
In recent years, unitary QCA have attracted attention as models for both 
specific quantum information processing tasks \cite{Macchiavello}
and as underlying models of emergent causal theories such as quantum field theory 
\cite{Ariano1,Ariano2,ArrighiDirac,ArrighiLorentz,Perinotti}  
and quantum gravity \cite{Lloyd,DSR}.
The model we adopt here, first introduced in \cite{papero1}, detaches itself from all the above mentioned 
approaches in that, while maintaining causality of interactions and translational invariance,
we will drop the unitarity requirement. 
The class of QCA we will consider is indeed defined within the more general formalism of CP-maps,
in which quantum noise is naturally a part of the picture and the dynamics is suitable to describe the evolution
of both open and close quantum systems. 
This framework then allows for a description of quantum state transfer processes in which the effects related to a certain
degree of unavoidable coupling between system and environment can be taken into account, 
which is something that has been very little explored so far in this context (with the notable exception 
of \cite{Chandra}; also see \cite{romanelli1,romanelli2}, where the whole quantum walk dynamics is unitary
and noise and mixedness are introduced by partial tracing over the position degrees of freedom).

In general, when dealing with spatially discrete quantum systems, the aim of 
the protocol is to transfer with unit fidelity an unknown arbitrary quantum state from a sender to a receiver 
through a network of interacting quantum systems, which is initially in its ground (or \apicisx vacuum\apicidx) state.
If qubits are considered, then a generic state prepared on the sender qubit $s$ ($\ket{\psi}_s=\alpha\ket{0}_s+\beta\ket{1}_s$) must be dynamically
transported through the network in a finite time $t^*$ to the receiver qubit $r$ (and possibly read out):
\begin{equation}
\ket{\psi}_s\otimes \ket{00\dots 0}_{network}\otimes \ket{0}_r
\quad \stackrel{t^*<\infty}{\longrightarrow} \quad \ket{\phi}\otimes \ket{\psi}_r  
\label{eq:banalita1}
\end{equation}
where the resulting state of network and sender $\ket{\phi}$ at the end of the process does not matter.    
Of course a sequence of unitary swap gates would be a straightforward solution, but it would require
a considerable amount of control over the dynamics as well as complete absence of noise.
Since we are interested in venturing into the noisy regime, Eq.(\ref{eq:banalita1}) must be recast into:
\begin{equation}
\ket{\psi}_s\otimes \ket{00\dots 0}_{network}\otimes \ket{0}_r
\quad \stackrel{t^*<\infty}{\longrightarrow} \quad \rho(t^*) \mbox{  s.t.  }
\rho_r(t^*)\equiv \hbox{Tr}_{\neg r}\left( \rho(t^*)\right)=\ketbra{\psi}{\psi},  
\label{eq:banalita2}
\end{equation}
where $Tr_{\neg r}(\rho)$ denotes partial tracing on the whole network but the receiver qubit $r$. 
This paper is devoted to studying the transport of the quantum state through qubit lattices 
governed by noisy QCA evolutions.

The plan of the paper is as follows.
In the next Section (Sec.\ref{sec:basicQCA}) the formal definition of QCA and
the equivalent operational structure is given.
In Sec.\ref{sec:model} the model is described by first introducing
a QCA dynamics on a linear chain of qubits suitable for energy transport,
defined in the single excitation sector of the Hilbert space.
By enlarging the Hilbert space, allowing for a global vacuum state,
a QCA capable of implementing quantum state transfer is obtained, and its features discussed.
The transport performance is investigated through numerical simulations
in Sec.\ref{sec:results} and, finally, in Sec.\ref{sec:conclusion} we summarize and conclude. 

%%%%%%%%%%%%%%%%%%%%%%%%%%%%%%%%%%%%%%%%%%%%%%%%%%%%%%%%%%%%%%%%%%%%%%% QCA BASICS
\section{Basic notions of QCA}\label{sec:basicQCA}
\begin{figure*}[t!]
\subfigure[]{\includegraphics[scale=0.5]{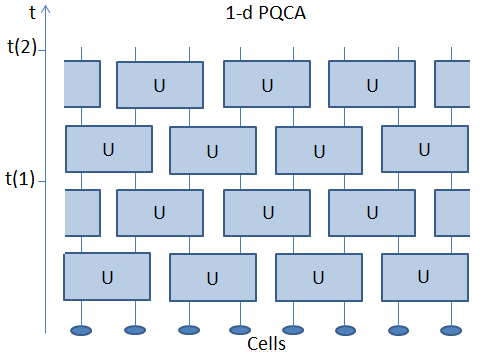}}
\subfigure[]{\includegraphics[scale=0.65]{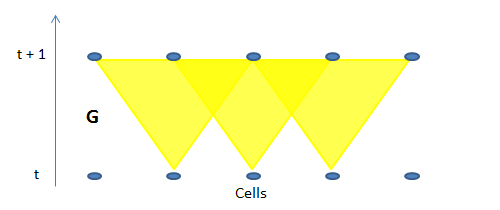}}
\caption{(a) Sketch of the 1-dimensional $1/2$ nearest neighbor QCA evolution in time. 
         Wires are cells, time flows upwards. One iteration of the automaton corresponds to
         the application of two layers of local two-qubit unitaries.
         (b) The light cone of a QCA.}
\label{fig:PQCAArrighi}
\end{figure*}

In order to justify the constructive approach we will follow later on to define noisy QCA, it is
beneficial to see how unitary QCA structures may emerge in an axiomatic framework.
Consider a QCA on an n-dimensional lattice and set a neighborhouring scheme for the local 
interactions.\footnote{The following definitions are taken from \cite{ArrighiQCA}. Note that in that work, the 
axiomatic definition of QCA involves the notion of \textit{quiescent states}. However, being it unnecessary
in this context, we will not introduce it.}
At each node of the lattice is placed a quantum
system (a qubit, for instance), whose internal states belong to a discrete finite set $\Sigma$.
All the possible \textit{finite} configurations $c \in \mathcal{C}^{\Sigma}$ the global state of the analogus classical
automaton can be in %\footnote{It is understood that $\mathcal{C}^{\Sigma}$ needs to be a \textit{finite} set.}
are now associated to an orthonormal basis $\{\ket{c}\}$ for the global
Hilbert space $\mathcal{H}_{\mathcal{C}^{\Sigma}}$ of the QCA. The state of the QCA 
at any time $t$ is a unit vector in $\mathcal{H}_{\mathcal{C}^{\Sigma}}$ and it can of course
be expressed as a superposition of the basis vectors $\{\ket{c}\}$.
The action of a QCA can be represented by a linear global map
$\mathcal{G} :\mathcal{B}(\mathcal{H}_{\mathcal{C}^{\Sigma}})\rightarrow \mathcal{B}(\mathcal{H}_{\mathcal{C}^{\Sigma}})$.
In order for a QCA to be axiomatically well defined,
the map $\G$ has to be unitary and shift invariant with respect to lattice translations.
In addition, $\G$ needs to satisfy the following definition of causality:  
\begin{defn}
\textbf{(Causality)} Let the global state of the QCA be identified with a trace-one positive operator $\rho$;
A linear map $\mathcal{G} :\mathcal{B}(\mathcal{H}_{\mathcal{C}^{\Sigma}})\rightarrow \mathcal{B}(\mathcal{H}_{\mathcal{C}^{\Sigma}})$  
is said to be causal iff for any $\rho,\rho'$ two states over $\mathcal{H}_{\mathcal{C}^{\Sigma}}$, and for any site $x $: 
\end{defn}
\begin{equation} \label{eq:causality}
\hbox{Tr}_{\mathcal{L}/\mathcal{N}_x}(\rho)=\hbox{Tr}_{\mathcal{L}/\mathcal{N}_x}(\rho') \quad\Rightarrow\quad 
\hbox{Tr}_{\mathcal{L}/x}\left[\G(\rho)\right]=
\hbox{Tr}_{\mathcal{L}/x}\left[\G(\rho')\right],
\end{equation}
where $\hbox{Tr}_{\mathcal{L}/\mathcal{N}_x(x)}$ means tracing over the entire lattice but the 
neighborhood $\mathcal{N}_x$(the site $x$).
In words, this definition means that to know the state of a site $x$ after one application of the automaton
we only need to know the state of
its neighborhood before the evolution.\footnote{This is a very strict notion of causality, 
in that it allows for a sharp definition of information light cones (Fig.\ref{fig:PQCAArrighi}(b)).
This contrasts with the exponentially vanishing instantaneous correlations between any two different regions of the lattice
allowed by any Hamiltonian continuous time evolution \cite{Lieb}.}
Hence:
\begin{defn}
\textbf{(Quantum Cellular Automata)}
An n-dimensional Quantum Cellular Automaton is an operator $G$ which is unitary, shift invariant and causal.
\end{defn}

For lattice dimension $d=1$, this axiomatic definition implies a precise operational structure \cite{ArrighiPQCA} 
by means of a $d=1$, $\frac{1}{2}$-nearest neighbor Partitioned QCA (PQCA). Because of that, in the following we will
be concerned with this kind of systems,
whose global evolution (depicted in Fig.\ref{fig:PQCAArrighi}(a)) can be expressed as:
\begin{equation}
\label{eq:1dPQCA}
\rho(t+1) = \G(\rho) = G\rho(t)G^{\dagger},\quad \mbox{with} \quad 
G = \left[ \sigma_{-1}\left( \bigotimes_{\mathcal{N}_x}U\right) \sigma_{+1}\left( \bigotimes_{\mathcal{N}_x}U\right) \right],
\end{equation}
where $U$ is the local unitary acting upon each neighborhood, $\sigma_{\pm 1}$ is the one-site right/left shift
of the lattice and $\mathcal{N}_x$ labels the $xth$ neighborhood. \\

The model we are going to describe in the next section (\ref{subsec:SES}) is based on this architecture,
but, as we will consider noise, we will not have, in general, a unitary scattering acting on a two-qubit Hilbert space (Fig.\ref{fig:PQCAArrighi}(a)).
We remark though that this will not be just a straightforward substitution ($U \rightarrow$ CP-map),
in that the class of local CP-maps we will introduce only acts on the one excitation subspace 
of a two-qubit neighborhood. As we shall see, our constructive approach will be dictated by 
specific arguments related to the possibility of embedding classical stochastic processes (Markov chains) 
within the more general, quantum, dynamics.
%%%%%%%%%%%%%%%
%\begin{figure}[htb!]
%\includegraphics[scale=0.65]{figures/lightcone1}
%\caption{The light cone of a QCA, as it is clear from Fig.\ref{fig:PQCAArrighi}.}
%\label{fig:QCAlightcone}
%\end{figure*} 
%%%%%%%%%%%%%%%

%%%%%%%%%%%%%%%%%%%%%%%%%%%%%%%%%%%%%%%%%%%%%%%%%%%%%%%%%%%%%%%%%%%%%%%% MODEL
\section{The model} \label{sec:model}
In this section, the model for quantum state transfer is described.
Following Ref.\cite{papero1}, the dynamics will be defined by first
considering the problem of transferring a single excitation. Once this is done, the necessary step in order
to be able to transfer a qubit will be to enlarge the Hilbert space of the dynamics such that a 
global vacuum state can be accommodated.    

%%%%%%%%%%%%%%%%%%%%%%%%%%%%%%%%%%%%%%%%%%%%%%%%%%%%%%%%%%%%%%%%%%%%%%%%  SINGLE EXC. DYNAMICS
\subsection{Single excitation sector dynamics} \label{subsec:SES}
Here we develop a formalism for describing transport of a single-particle excitation along a linear chain of qubits.
The main aim is to to establish a new framework in which a quantum and a stochastic classical dynamics can be simply and fairly compared.
To do so, as a starting point we consider the simplest and most general stochastic process: 
A random walk with fixed transition probabilities in the forward and backward directions along the chain. 
We then embed the classical transition probabilities defining the random walk within the formalism of quantum operations
and give the global evolution a QCA structure, finding a dynamics that allows to explore the transition
between purely classical and purely quantum transport. \\

Consider a classical random walk between two sites. This is the prototipical Markov process,
represented by a general stochastic transition matrix of the form
\be
T_{p,q} = 
\begin{bmatrix}
1-p & q \\
p & 1-q
\end{bmatrix}, \quad {\rm with} \quad 0 \le p,q \le 1 \,  \label{stocco}
\ee
acting on dichotomic probability distribution vectors of the kind $v_m=(m,1-m)^{\sf T}$, with $0\le m \le 1$. 
In the random walk picture, 
the parameters $p$ and $q$ represent the probabilities to jump from one site to the other at each 
discrete time step of the evolution. When there is no bias between the two directions of the walk,
i.e. $p=q$, the transition matrix $T_{p,p}$ is called doubly stochastic. 
This classical dynamics may be related to a quantum dynamics if one takes into account a specific class
$\mathcal{C}$ of completely positive (CP) maps that, once a privileged basis is set,
send diagonal density operators into diagonal density operators.\footnote{Quite significantly, 
another class of such maps has been recently adopted to establish a resource theory of quantum 
coherence \cite{baumgratz}.} 
It is possible to find a simple construction which allows one to reproduce any possible two-dimensional stochastic matrix 
by considering a subset of ${\mathcal C}$ acting on a one-qubit system.

First, note that one-qubit diagonal density matrices can be trivially bijectively mapped into dichotomic probability 
distribution vectors, as per $B:\rho_m={\rm diag}(m,1-m)\mapsto v_m=(m,1-m)^{\sf T}$, where we have denoted
such a bijection by $B$. 
We will conventionally refer to the parameter $m$ as the probability of populating the excited state of the qubit,
or \apicisx excitation probability\apicidx.
Two different kinds of quantum noise, ultimately selecting the classical basis,
will enter the picture of our dynamics. The dephasing channel $\Phi_{\xi}$, which arguably represents the
most natural decoherence mechanism in several practical cases, has the effect of damping all coherences
in a density matrix by a factor $\sqrt{1-\xi}$, while leaving the diagonal elements untouched.   
Its Kraus operators are:
\begin{small}\be
D_0=\sqrt{1-\xi}\id,\;
D_1=\sqrt{\xi}(\id+\sigma_z)/2 , \;
D_2=\sqrt{\xi}(\sigma_z-\id)/2 \, , \label{dephase}
\ee \end{small}
\noindent where the dephasing strength parameter $\xi$ is such that $0\le\xi\le1$,
and $\sigma_j$ for $j=x,y,z$ stand for the Pauli matrices.
We shall refer to the CP-map $\Phi_1$, whose effect is setting to zero all the off-diagonal elements, as
\apicisx total\apicisx, or \apicisx complete\apicidx dephasing. 
The amplitude damping channel $\Xi_{\eta}$, is defined by the Kraus operators 
\begin{small}
\be
L_{0,\eta} = (\id+\sigma_z)/2 + \sqrt{1-\eta}(\id-\sigma_z)/2 , \; L_{1,\eta}=\sqrt{\eta}(\sigma_x+i\sigma_y)/2 \, , \label{ampli}
\ee
\end{small}
\noindent with $0\le\eta\le 1$. As the dephasing, it damps all coherences by a factor $\sqrt{1-\eta}$, while also affecting
the populations on the diagonal. It will be convenient to introduce the \apicisx swapped\apicidx amplitude damping channel,
the extention of the definition of $\Xi_{\eta}$ to negative $\eta$ ($-1\le\eta\le0$), characterized by the Kraus operators
$\sigma_x L_{0,|\eta|}\sigma_x$ and $\sigma_x L_{1,|\eta|} \sigma_x$. 

We can now show two statements relating classical stochastic maps and single qubit dynamics:  
\begin{prop}
Any two-dimensional doubly stochastic map may be represented on diagonal density matrices by the action of a unitary map followed by complete dephasing.
\label{prop:prop1}
\end{prop}

{\em Proof:}
Let $U_{\theta,\gr{\varphi}}$ be a generic $2\times2$ unitary matrix parametrized as  
\be
U_{\theta,\gr{\varphi}} =
\begin{bmatrix}
\cos\theta & \sin\theta {\rm e}^{i\varphi_2} \\
-\sin\theta{\rm e}^{i\varphi_1} & \cos\theta {\rm e}^{i(\varphi_1+\varphi_2)}
\end{bmatrix}
\label{eq:genU}
\ee
with $0\le\theta\le\pi$, $\gr{\varphi}=(\varphi_1,\varphi_2)$ and $0\le\varphi_1,\varphi_2\le2\pi$.
The action of the map $B\circ \Phi_1\circ U_{\theta,\gr{\varphi}}$ on a diagonal density matrix $\rho_m=\mbox{diag}(m,1-m)$ (where $U_{\theta,\gr{\varphi}}$ is understood to act by similarity)
is analogous to the action of the stochastic map $T_{p,p}$ of Eq.(\ref{stocco}) on the
probability vector $v_m=(m,1-m)^{\sf T}$: 
\be
B\left(\Phi_1\left(U_{\theta,\gr{\varphi}} \rho_m U_{\theta,\gr{\varphi}}^{\dagger} \right)\right) = 
T_{\sin^2\theta,\sin^2\theta} \gr{v}_m \; ,
\ee
upon identifying $p=\sin(\theta)^2$. 
It is thus always possible to reproduce any doubly stochastic maps through a proper choice of $\theta$. $\square$ \\
\begin{prop}
Any two-dimensional stochastic map may be represented on diagonal density matrices
by the action of a completely dephased unitary map, followed by an amplitude damping channel.
\label{prop:prop2}
\end{prop}

{\em Proof:}
By direct application of the map $\Xi_{\eta} \circ \Phi_1\circ U_{\theta,\gr{\varphi}}$ we find
a state $\rho_{m'}$ with excitation probability:
\be
m' = c(1-|\eta|)m + \frac{1+|\eta| c -c +\eta}{2} \; , \label{muno1}
\ee
while the action of $T_{p,q}$ on the vector $v_m$ gives a vector $v_{m'}$,
with $m' = (1-p-q)m+q$. By equating the two new excitation probabilities one obtains the desired 
relationships between the parameters defining the two different maps:
\be
\eta=q-p \; ; \qquad \cos(2\theta)=\frac{1-p-q}{1-|q-p|} \; .
\ee
Even though we are not claiming this is the only possible or the most general embedding,
it is clear that all values of $p$ and $q$, and thus all two-dimensional stochastic maps,
can be reproduced by the open dynamics we considered by 
an appropriate choice of $\eta$ and $\theta$.
Proposition \ref{prop:prop2} (and hence \ref{prop:prop1}) can then be translated into:
\be
B\left[\Xi_{|q-p|}\left(\Phi_1\left(U_{f(\theta),\gr{\varphi}} \rho_m 
U_{f(\theta),\gr{\varphi}}^{\dag} \right)\right)\right] = 
T_{p,q} v_m \; ,
\label{eq:ourQCA}
\ee
where 
\be 
f(\theta):=\frac12\arccos\left(\frac{1-p-q}{1-|q-p|}\right).
\label{eq:arcos}
\ee
$\square$ \\ 

Using this embedding of two-dimensional stochastic maps into dissipative qubit dynamics,
it is now possible to define an extended single excitation dynamics on a 1-d lattice of
qubits employing a partitioned quantum cellular automata structure.  
Let us consider the case of a single excitation transfer from one end to the other
of a chain composed of $N$ qubits.
As it is customary when dealing with excitation transport processes, we shall restrict to
the single excitation subspace of the global, $2^N$-dimensional Hilbert space. This assumption,
which dramatically simplifies the treatise,
is for instance biologically well justified when modeling photosynthetic systems, which host only
one excitation at a time during a transport dynamics.
In this sector, the Hilbert space (whose dimension is now equal to $N$)
is spanned by the basis $\{\ket{n}\}$ ($1\le n \le N$), with $\ket{n}$ representing the state
in which the $n$-th qubit is in an excited state, while all the others are in the ground state.   
We can now define a CP map $\Omega^{(n)}_{\eta,\xi,\theta,\gr{\varphi}}$ as the composition
of a unitary, a phase damping and an amplitude damping acting locally, i.e. on a two dimensional
subspace spanned by $\ket{n}$ and $\ket{n+1}$:
$
\Omega^{(n)}_{\eta,\xi,\theta,\gr{\varphi}} (\rho) = \Xi^{(n)}_{\eta}\left(\Phi^{(n)}_{\xi}\left(U^{(n)}_{\theta,\gr{\varphi}}\rho 
U^{(n)\dag}_{\theta,\gr{\varphi}}\right)\right).
$ 
Recalling the scheme described in Sec.\ref{sec:basicQCA}, the repetition of an application of
a layer of $\Omega^{(n)}_{\eta,\xi,\theta,\gr{\varphi}}$ on disjoint blocks of $2$ qubits
(neighborhoods) followed by the same operation
shifted by one site on the lattice gives a quantum cellular automaton, i.e., a discrete, causal and translationally invariant 
quantum evolution on a lattice. Hence, the global map
\cite{ArrighiQCA,Schumacher,ArrighiPQCA,arrighigrattage}
\be
\Omega_{\eta,\xi,\theta,\gr{\varphi}} = 
\bigotimes_{l\,{\rm odd}}\Omega^{(l)}_{\eta,\xi,\theta,\gr{\varphi}} \bigotimes_{l\, {\rm even}}\Omega^{(l)}_{\eta,\xi,\theta,\gr{\varphi}} \;
\label{eq:QCAmap}
\ee
%%%%%%%%%%%%%% 
\begin{figure*}[t!]
\includegraphics[scale=0.3]{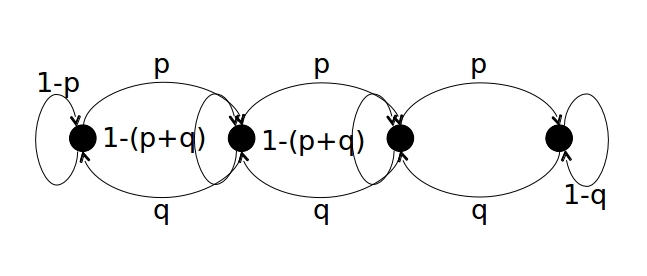}
\includegraphics[scale=0.3]{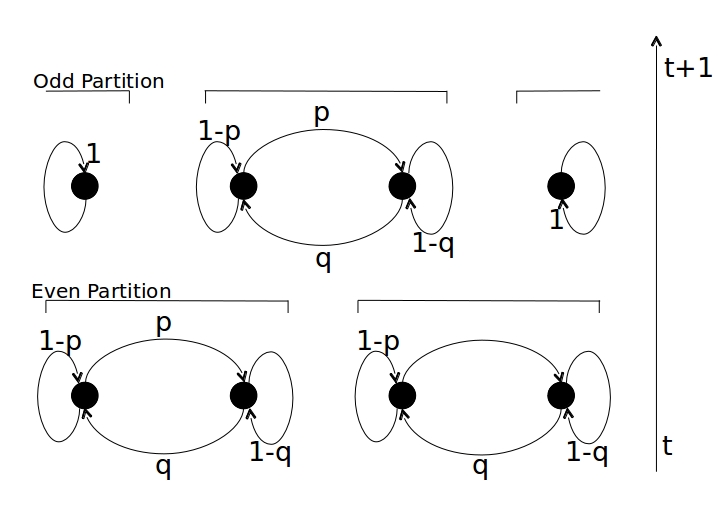}
\caption{(\textit{left}) The classical random walk on a chain of $N=4$ sites; (\textit{right})
          One time step scheme of the equivalent partitioned random walk, which represents the
          classical dynamics QCA (\ref{eq:QCAmap}) is reduced to when total dephasing ($\xi=1$) is applied.}
\label{fig:RW}
\end{figure*}
%%%%%%%%%%%%%%
defines a class of quantum cellular automata\footnote{Giving an axiomatic definition
to open QCA have proven to be difficult, if not impossible \cite{ArrighiProbQCA},
so maintaining the denomination of quantum cellular automata is in a sense not completely rigorous.
Nonetheless, the maps we have constructed fulfill the same requirements of causality
and (block-)translational invariance which are the peculiar characteristics defining QCA.}
 in which it is possible to range between probabilistic
classical dynamics and quantum dynamics only by means of tuning one noise parameter,
the dephasing strenght $\xi$, which drives the decoherence mechanism in any open quantum systems.
When $\xi=1$, a Markov chain classical transfer takes place ((Fig.\ref{fig:RW})),
while decreasing the dephasing parameter an open (noisy) quantum dynamics ($0<\xi<1$) 
and a closed (unitary) one ($\xi=0$; $\eta=0$) are realized.
Therefore in this framework, once the local transition probabilities $p$ and $q$ are set,
the quantum to classical transition can be explored in a meaningful way, as 
the difference between the two regimes is truly only down to quantum coherence.   

Our model was applied in \cite{papero1} to the study of energy excitation transfer through the lattice 
by comparing, at given local transition probabilities, the performance of a classical process with that of quantum dynamics where 
coherent phases are allowed to develop and interfere along the chain.
Along with highlighting differences in performance between the two regimes,
the model prooved to succeed in capturing emerging purely quantum distinctive features
such as noise-assisted transport and destructive interference. %(see e.g. \cite{plenio}).  

%%%%%%%%%%%%%%%%%%%%%%%%%%%%%%%%%%%%%%%%%%%%%%%%%%%%%%%%%%%%%%%%%%%%%%%%%%%%%%%%% EXTENDED DYNAMICS
\subsection{Enlarging the Hilbert space}
While in the previous subsection we defined a framework which was sufficient to study single excitation 
transport, our aim now is to apply the same dynamics to the transport of a quantum state,
i.e. we want to investigate how well this novel noisy QCA dynamics can transport initial 
on-site coherences, besides
excitations.
In order to do that, i.e. to transport the state of a qubit (as in Eqs.(\ref{eq:banalita1}) or (\ref{eq:banalita2})),
we need to extend the action of the automata CP maps
from the single excitation sector (SES), to a Hilbert space which must also include
a global vacuum state, the state in which there are no excitations in the chain.
Let us recall that the full Hilbert space $\mathcal{H}^{\otimes N}$ (where $\mathcal{H}$ indicates the 
Hilbert space of a qubit) can always be decomposed into the direct sum of
its number conserving sectors:
\be
\mathcal{H}^{\otimes N}=\bigoplus_{e=0}^{N}\mathcal{H}_e \; .
\ee
With the assumption that no other sectors are populated,
our maps will then be defined in the subspace with $0$ and $1$ excitations, $\mathcal{H}_0 \oplus \mathcal{H}_1$. 
Let us stress that there is a systematic way of going from this reduced subspace to
the full $\mathcal{H}^{\otimes N}$, making the notion of partial trace (and thus reduced state)
still meaningful and well defined (see Appendix \ref{appendixA}).

In terms of the operators involved in the evolution of the system, including a global vacuum
simply translates into adding a row and a column to the global density matrix $\rho$
and to the Kraus operators composing the global channel. Let us then focus on the Kraus
operators formalism representing the map in Eq.(\ref{eq:QCAmap}). 

First, using Eq.(\ref{eq:genU}) and Eq.(\ref{eq:arcos}) we can express $U_{\theta,\gr{\varphi}}$ explicitely
in terms of the transition probabilities $p$ and $q$ 
\be
U^{(l)}=\frac{1}{\sqrt{1-\eta}}
\begin{bmatrix}
\sqrt{1-p} & \sqrt{q}e^{i\phi_2} \\
\sqrt{q}e^{i\phi_1} & -\sqrt{1-p}e^{i(\phi_1+\phi_2)} \\ 
\end{bmatrix}.
\ee 
From now on we will be regarding $p$ and $q$ as the probability of the classical stochastic
dynamics to jump one site on the left and on the right, respectively. We will always consider
the case $p\ge q$, as we also arbitrarily choose to set the transfer direction from left to right.
The superscript $(l)$ will be used throughout the rest of the paper to indicate local operators.
As regards the general expression of the local channel representing the map 
$\Omega^{(n)}_{\eta,\xi,\theta,\gr{\varphi}}$, we find:
\begin{small}
\be
K_0^{(l)}=\sqrt{\frac{1-\xi/2}{1-\eta}}
\begin{bmatrix}
\sqrt{1-\eta}\sqrt{1-p} & \sqrt{1-\eta}\sqrt{q}e^{i\phi_2} \\
\sqrt{q}e^{i\phi_1} & -\sqrt{1-p}e^{i(\phi_1+\phi_2)} \\
\end{bmatrix},
\label{eq:k0} 
\ee
\end{small}
\begin{small}
\be
K_1^{(l)}=\sqrt{\frac{\xi/2}{1-\eta}}
\begin{bmatrix}
\sqrt{1-\eta}\sqrt{1-p} & \sqrt{1-\eta}\sqrt{q}e^{i\phi_2} \\
-\sqrt{q}e^{i\phi_1} & \sqrt{1-p}e^{i(\phi_1+\phi_2)} \\
\end{bmatrix}, \\
\label{eq:k1}
\ee
\end{small}
\begin{small}
\be
K_2^{(l)}\sqrt{\frac{\eta}{1-\eta}}=
\begin{bmatrix}
0 & 0 \\
\sqrt{1-p} & \sqrt{q}e^{i\phi_2} \\
\end{bmatrix}.
\label{eq:k2}
\ee
\end{small}
The above operators are obtained by changing to a unitary equivalent representation 
for the dephasing channel (\ref{dephase}) composed of only two Kraus operators \cite{nielsen} written in terms of
a rescaled phase damping parameter $\xi'=1-\xi/2$, and then by multiplying $U^{(l)}$ by the combination of the 
application of the new dephasing channel followed by the amplitude damping channel (Eq.(\ref{eq:ourQCA})).  

Given the QCA structure, the global channel in the SES is identified by three $N\times N$ block diagonal
matrices $\{{K_{\mu}^{(s)}}\}$ (the superscript $(s)$ identifies operators acting on the SES)
in which the blocks are the corresponding local operators $\{K^{(l)}_{\mu}\}$.
In the full (SES plus vacuum) Hilbert space $\tilde{\mathcal{H}}$ we can 
write\footnote{Assuming the operators in Eq.(\ref{eq:globalK}) refer to the even partition of the lattice,
the odd partition's ones will simply be:
\be
\tilde{K_{\mu}}=
\begin{bmatrix}
z_{\mu} & \gr{V}^{\dagger}_{\mu} \\
\gr{W}_{\mu} & \sigma K_{\mu}^{(s)}\sigma^{\dagger} \\
\end{bmatrix}, \quad \mu=(0,1,2),
\label{eq:globalKodd}
\ee
where $\sigma$ is the one site shift operator. Some extra care must be taken 
when applying the shift to chains with no periodic boundary conditions (see later on).} 
\be
\tilde{K_{\mu}}=
\begin{bmatrix}
z_{\mu} & \gr{V}^{\dagger}_{\mu} \\
\gr{W}_{\mu} & K_{\mu}^{(s)} \\
\end{bmatrix}, \quad \mu=(0,1,2),
\label{eq:globalK}
\ee
where $\{\gr{V}_{\mu}\}$ and $\{\gr{W}_{\mu}\}$ are $N$-dimensional vectors,
$\{z_{\mu}\}$ are the $\ket{0}\bra{0}$ scalar entries in the enlarged Hilbert space,
and we have introduced a {\em tilde}-notation such that $\tilde{X}$ represents an $(N+1)$-dimensional operator.
One time step of the QCA evolution of the system [Eqs.(\ref{eq:QCAmap})]
can then be explicitely written as:
\be
\tilde{\Omega}(\tilde{\rho}) = \sum_{\nu=0}^{2}\tilde{K}_{\nu}^{(odd)}
\left[\sum_{\mu=0}^{2}\tilde{K}_{\mu}^{(even)}\tilde{\rho}\tilde{K}_{\mu}^{(even)\dagger}\right]\tilde{K}_{\nu}^{(odd)\dagger} 
\ee

The condition for the new Kraus operators to sum up to the identity
$\sum_{\mu}\tilde{K_{\mu}}^{\dagger}\tilde{K_{\mu}}=\id$ now reads:
\begin{small}
\be
\begin{bmatrix}
|z_{\mu}|^2+||\gr{W}_{\mu}||_2^2 & c.c. \\
(z_{\mu}\gr{V}_{\mu}+{K_{\mu}^{(s)}}^{\dagger}\cdot \gr{W}_{\mu}) & (\gr{V}_{\mu}\cdot \gr{V}_{\mu}^{\dagger}+{K_{\mu}^{(s)}}^{\dagger}\cdot K_{\mu}^{(s)}) \\ 
\end{bmatrix} = \id,
\label{eq:id}
\ee
\end{small}
which leads to a set of constraints:
\be
\label{eq:constraints}
\left\{
\begin{array}{lr}
\sum_{\mu}(|z_{\mu}|^2+||\gr{W}_{\mu}||_2^2)=1 & \\
\sum_{\mu}{K_{\mu}^{(s)}}^{\dagger}\cdot \gr{W}_{\mu}=0 & \\
\gr{V}_{\mu}=0, \quad \forall \mu & \\
\end{array}
\right.
\ee 
where $\gr{V}_{\mu}=0$, $\forall \mu$, is due to the fact that the outer product
of a vector with itself gives a nonnegative matrix, and we already had
$\sum_{\mu}K^{(s)\dagger}_{\mu}K^{(s)}_{\mu}=\id$.   

Looking at Eq.(\ref{eq:globalK}), we see that the two vectors $\gr{W}_{\mu}$ and $\gr{V}_{\mu}$
can be thought of, respectively, as excitation \apicisx pumping\apicidx
and dissipation processes, in that
the former represents the entries 
of $\{\tilde{K}_{\mu}\}$ that account for the creation of an excitation from the vacuum
($\{\gr{W}_{\mu}\}\leftrightarrow \{(\ket{n}\bra{0})_{\mu}\}_{\mu=1}^{N}$),
while the latter provides those entries such that excitations are 
annihilated into the vacuum
($\{\gr{V}_{\mu}\}\leftrightarrow \{(\ket{0}\bra{n})_{\mu}\}_{\mu=1}^{N}$).
Notice that we do not allow for more than one excitation in the lattice during the transport
and thus the role of the $\{\gr{W}_{\mu}\}$ vectors in the dynamics is to increase at each time step of
the evolution the probability of having the excitation somewhere in the chain.
Nonetheless, the two sets of vectors are in a broad sense representative of
dissipation and excitation effects when hypothesizing a physical realization of a QCA 
quantum state transport device. 
  
Given the Kraus operators of the single excitation sector, $K_{\mu}^{(s)}$, Eqs.(\ref{eq:constraints})
leave one with a certain freedom in the choice of the $\gr{W}_{\mu}$.
In other words, for each noisy QCA dynamics in the SES, we are left with a whole class of dynamics
in the extended Hilbert space. 
We notice that one can take advantage of the peculiar block structure of the QCA dynamics [Eq.(\ref{eq:QCAmap})]
in order to characterize the  vectors $\gr{W}_{\mu}$.
For closed chains with $N$ even, each of the two QCA partitions are formed of 
$N/2$ blocks of local ($2\times 2$) Kraus operators lying on the diagonal of the global operator.
Looking back at the set of constraints (Eq.(\ref{eq:constraints})), the aforementioned block structure 
implies that $\sum_{\mu}{K_{\mu}^{(s)}}^{\dagger}\cdot \gr{W}_{\mu}=0$ can be regarded as a local condition,
as it can be rewritten in the form: 
\be
\label{eq:localT}
\sum_{\mu}{K_{\mu}^{(l)}}^{\dagger}\cdot \gr{w}_{\mu}(\mathcal{N}_i)=0,
\quad \forall i,
\ee
where the vectors $\gr{w}_{\mu}(\mathcal{N}_i)$ are $N/2$ $2$-dimensional
local vectors %\footnote{$N/2-1$ for the odd partition without PBE.}
defined at each neighborhood $\{\mathcal{N}_i\}_{i=1}^{N/2}$.
These vectors can in principle be different from each other, but since we want to preserve
translational invariance (see Appendix \ref{appendixB}), we set them to be the same 
(for a given superoperator $\mu$) in every neighborhood, thus
dropping the neighborhood index dependence $\mathcal{N}_i$:
$ \gr{W}_{\mu}=\bigoplus \gr{w}_{\mu}$, with $\gr{w}_{\mu}=(w_{\mu}^0,w_{\mu}^1)$, $\forall \mu$.
This way this source of noise represents another homogeneous property of the
dynamics.

The introduction of a global vacuum, though,
implies that our extended interactions may not fulfill the stringent causality definition of Eq.(\ref{eq:causality}),
which is a crucial property for a QCA evolution.
Indeed, it turns out that additional constraints need to be added to the dynamics.
The question of which of our extended CP-maps are still causal is settled by the following statement:
\begin{prop} \label{prop:causalclass}
The class of proper - i.e. causal and translational invariant - noisy quantum cellular automata on 
a 1-dimensional qubit lattice, when the dynamics is restricted to the $0\oplus1$ sector of the
Hilbert space, is defined by the set of constraints: 
\be
\label{eq:allconstraints}
\left\{
\begin{array}{lr}
\gr{W}_{\mu}=\bigoplus^{N/2}\gr{w}_{\mu} \; , & \forall \mu \\
\gr{V}_{\mu}=0 \; , & \forall \mu \\
\sum_{\mu}(|z_{\mu}|^2+\frac{N}{2}||\gr{w}_{\mu}||_2^2)=1 & \\
\sum_{\mu}{K_{\mu}^{(l)}}^{\dagger}\cdot \gr{w}_{\mu}=0 & \\
\sum_{\mu}z_{\mu}w_{\mu}^i=\sum_{\mu}w_{\mu}^i=0 \; , & i=0,1 \\
\end{array}
\right.
\ee
\end{prop}

{\em Proof:} See Appendix \ref{appendixC}. \\

As a concluding remark for this section, we note that the fact that $\{\gr{V}_{\mu}\}$ (Eqs.(\ref{eq:allconstraints})) 
are null means that the probability of having no excitations in the chain cannot increase during the evolution.
Hence, the system is not subject to dissipation-like effects.

Of course one can in principle obtain other CP-maps in the enlarged Hilbert space, possibly with $\gr{V}_{\mu}\neq0$,
by rescaling the SES process (e.g. $K_{\mu}^{(s)}$ $\rightarrow$ $cK_{\mu}^{(s)}$, $\forall \mu$,
with $|c|<1$). However, we are interested in preserving the SES dynamics introduced in the previous
subsection, so we will only deal with the case characterized by Eqs.(\ref{eq:constraints}).

%%%%%%%%%%%%%%%%%%%%%%%%%%%%%%%%%%%%%%%%%%%%%%%%%%%%%%%%%%%%%%%%%%%%%%%%%%%%%%%%% NUMERICS
\section{Quantum state transfer} \label{sec:results}
We will consider two different settings: an open linear chain\footnote{Notice that in the absence
of periodic boundary conditions one drops \textit{global} translational invariance.} and a ring composed of
an even number $N$ of qubits.
In both cases we will assume the transfer of the quantum state to take place from left to right
($p\ge q$), from the first site to the antipodal one, i.e. site $N$ for a chain and $N/2+1$
for a ring.  
To quantify the transport, we will be employing the Uhlman fidelity between quantum states $\sigma$ an $\rho$: 
$F(\sigma,\rho)=Tr(\sqrt{\sigma}\rho\sqrt{\sigma})$. Since the initial state to be 
trasfer will be pure, in our case the fidelity is reduced to the form
$F(t)=\bra{\psi(0)}Tr_{\mathcal{L}/x}(\rho(t))\ket{\psi(0)}$, where the reduced state of the
target $x$ qubit is obtained by tracing the global state $\rho$ over the entire lattice
$\mathcal{L}$ but the qubit itself (see Appendix \ref{appendixA}). 
For any given setting, we will be considering as our figure of merit the average of the
fidelities obtained by simulating the transfer of an ensamble of initial states
drawn at random according to the Haar measure. \\

By observing the form of the local Kraus operators composing the channel,
Eqs.(\ref{eq:k0},\ref{eq:k1},\ref{eq:k2}), the first thing we note is that the closest the dynamics  
to unitarity, the more it is driven by the action of $K_0^{(l)}$; when the 
dynamics is purely unitary ($\xi=0$, $\eta=0$) $K_1^{(l)}$ and $K_2^{(l)}$ vanish.
Since we are investigating a quantum dynamics, we are mostly interested in both the unitary
and an \textquotedblleft almost unitary\textquotedblright regimes, the study of the latter being
relevant when envisaging a physical device which would naturally be exposed to some noise due
to unavoidable coupling with the environment. 
Thus, when in this two regimes, it seems appropriate to priviledge a tuple of the form
$\{c_{\mu}\}=(a,b,c)$, $a\gg b\simeq c$, for the relative weights of the global channel's
$\{z_{\mu}\}$ of Eq.(\ref{eq:globalK}).
%$\{\tilde{K}_{\mu}^{00}\}$ entries. 
Moreover, it is reasonable to expect that with such $\{c_{\mu}\}$ the transfer fidelity
will be higher. The same reasoning can be applied to the single excitation sector of the dynamics
when dealing with the \textit{odd} partition of an open chain.
In that case, the $N/2 - 1$ local operators do not act on the first and last qubits, so that 
the first and last entries of the global
operators of the \textit{odd} partition are just scalars which have to be set such that 
they sum up to one to meet the identity condition for the Kraus operators.
In this case we decide to set
$\tilde{K}_0^{11}=\tilde{K}_0^{NN}=1;$ $\tilde{K}_i^{11}=\tilde{K}_i^{NN}=0,$ for $i\neq 0$;
for the global dynamics we set $\{c_{\mu}\}=(1,0,0)$. \\

\begin{figure}[htb!]
\includegraphics[scale=0.66]{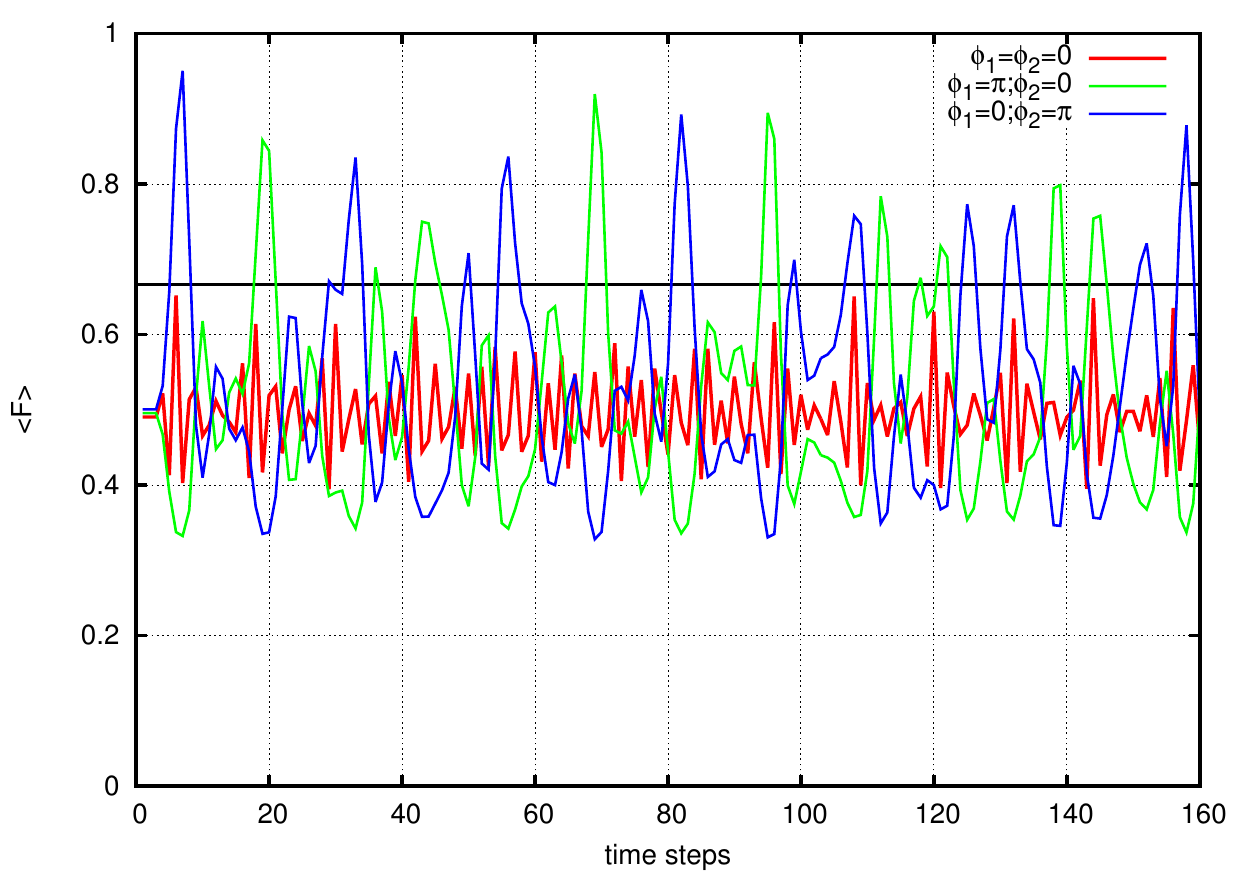}
\caption{Average fidelity $\langle F(t)\rangle$ over a sample of 1000 evolutions starting from initial states 
drawn at random from the Haar measure, for $p=q=0.5$ and three different choices of $(\phi_1,\phi_2)$
on a linear chain of $N=8$ qubits. No noise applied.
The straight horizontal line at $F=2/3$ shows the highest fidelity for classical transmission
of a quantum state.}
\label{fig:phases}
\end{figure}

Let us start analysing our numerical results by focusing on a linear open chain. 
In Fig.\ref{fig:phases} we show the average fidelity of transfer on a chain of
$N=8$ qubits for a fully unitary dynamics (i.e. $\eta=0$, $\xi=0$ and $\gr{W}_{\mu}=0$, $\forall \mu$)
with $p=q=0.5$. The three lines plotted correspond to three different choices for
the two free phases appearing in the general parametrization of $U^{(l)}$ (Eq.(\ref{eq:genU})).
%%%%%%%%%%%%%%%%%%%%%%%%%%%%%%%%%%%%%%%%%%
%%%%%%%%%%%%%%%%%%%%%%%%%%%%%%%%%%%%%%%%%%
As it can be clearly seen, these phases do play a role in the transferring process,
in that for both phases set to zero $\langle F(t)\rangle$ fluctuates around $\sim0.5$
(which is merely the average fidelity between two random Haar distributed states), 
whereas when either $(\phi_1,\phi_2)=(0,\pi)$ or $(\phi_1,\phi_2)=(\pi,0)$
the amplitude of the average fidelity fluctuations in time are much wider and there are
several times at which the fidelity is considerably higher than $2/3$ \cite{Fclassic}, which is the maximum value
attainable using only classical transmission of information ({\em measure and prepare} strategies).  
%%%%%%%%%%%%%%%%%%%%%%%%%%%%%%%%%%%%%%%%%%
\begin{figure*}[t!]
\subfigure[]{\includegraphics[scale=0.66]{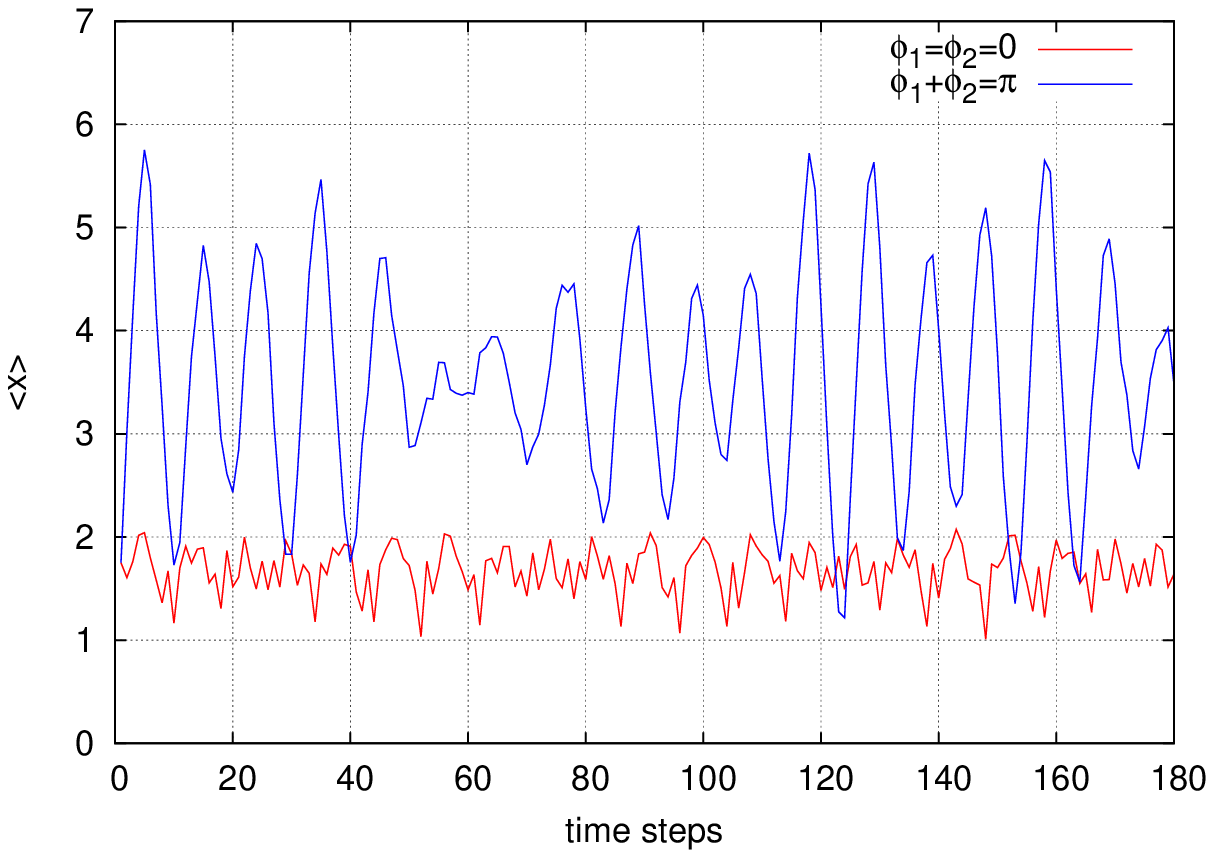}}
\subfigure[]{\includegraphics[scale=0.66]{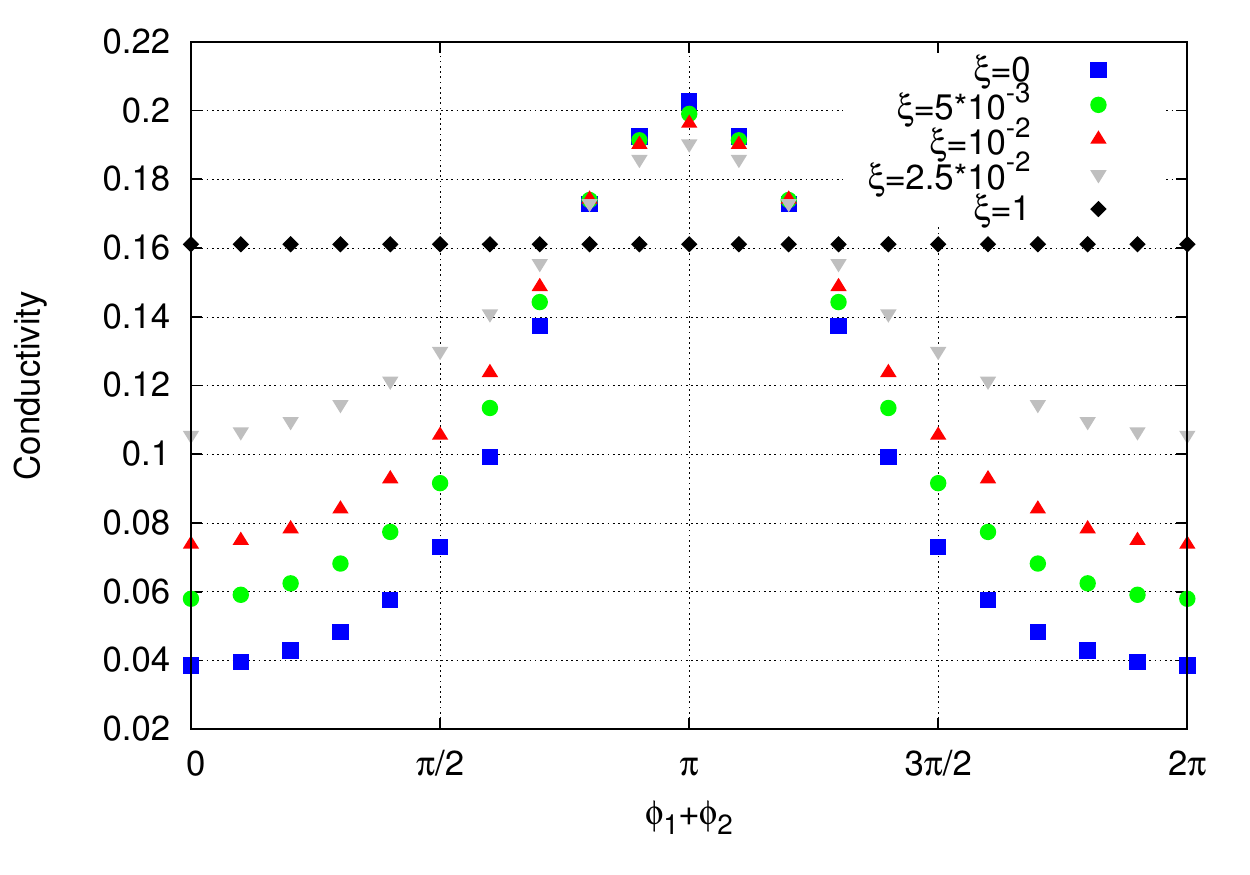}}
\caption{(a) Mean excitation position in time along a chain of $N=6$ qubits and $p=q=0.5$
for a dynamics in the sigle excitation sector. When $\phi_1+\phi_2=0$ a localization effect between
the first two sites of the chain takes place. (b) Excitation transport performance 
measured by the channel's conductivity $C(t_{end})=(1/t_{end})\sum_{t=1}^{t_{end}}\rho_{NN}(t)$ to the last site
at the end of the evolution ($t_{end}=160$) vs $\phi_1+\phi_2$ (same setting as in (a)).
When no dephasing is present (blue squares), by tuning the sum of the two phases
$\phi_1+\phi_2$ the transport can be highly suppressed due to the observed localization effect.
Interestingly enough, the introduction of some dephasing relaxes the localization, 
thus enhancing the transfer performance when $\phi_1+\phi_2$
is not close to the optimal value of $\pi$. When total dephasing $\xi=1$ is applied, the system 
undergoes a classical stochastic dynamics and the two phases cease to play any role (black diamonds)}
\label{fig:localization}
\end{figure*}
%%%%%%%%%%%%%%%%%%%%%%%%%%%%%%%%%%%%%%%%%%

This phase dependence could be related to an analogue effect observed when dealing
with the single excitation transport dynamics.
By tuning the sum of the two phases, in the SES an interesting localization phenomenon
takes place for which the probability of finding the excitation can be highly concentrated
between the first two sites of the chain at all times of the evolution (Fig.\ref{fig:localization}(a)).     
This of course affects negatively the transport performance, as shown in Fig.\ref{fig:localization}(b).
A detailed analysis of a similar localisation effect, possibly related to ours,
in a modelization of some biological systems, can be found in \cite{plenio}.
The analogy is strenghtened by the fact that in \cite{plenio}, like in our model, there is a phase
parameter that plays the same role in transport suppression. 
In the rest of the article we will be showing results obtained with the optimized choice of phases
$(\phi_1,\phi_2)=(0,\pi)$, unless explicitely stated. \\

\begin{figure*}[t!]
\subfigure[]{\includegraphics[scale=0.66]{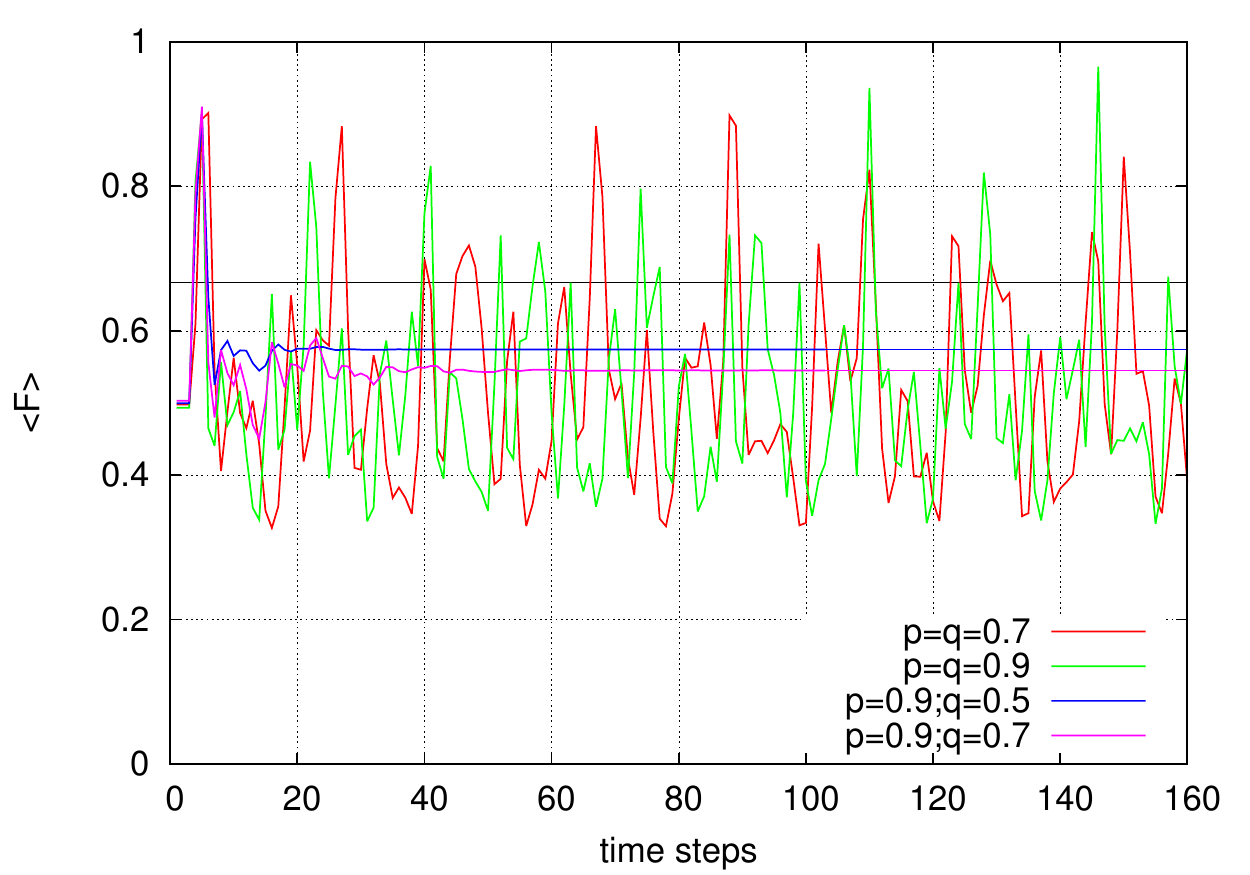}}
\subfigure[]{\includegraphics[scale=0.66]{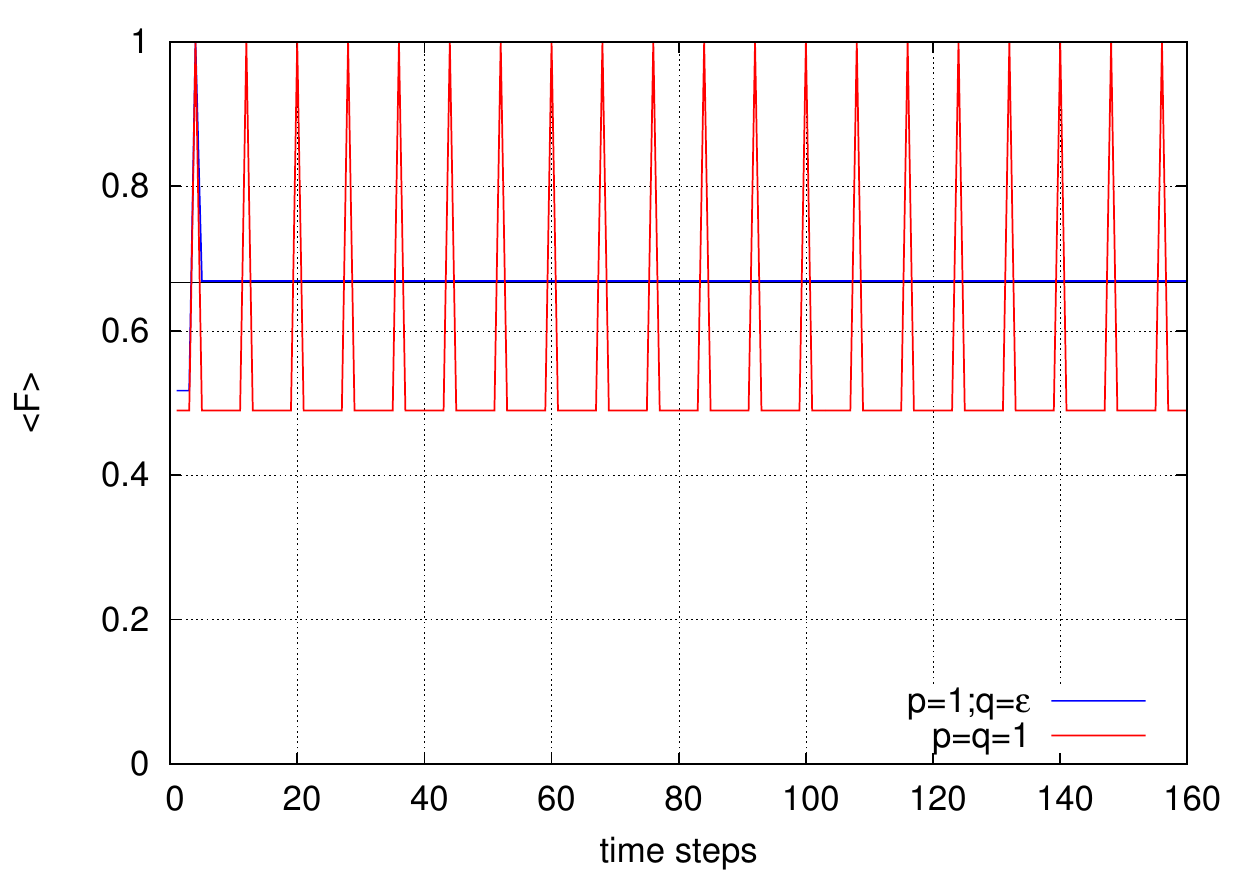}}

\subfigure[]{\includegraphics[scale=0.66]{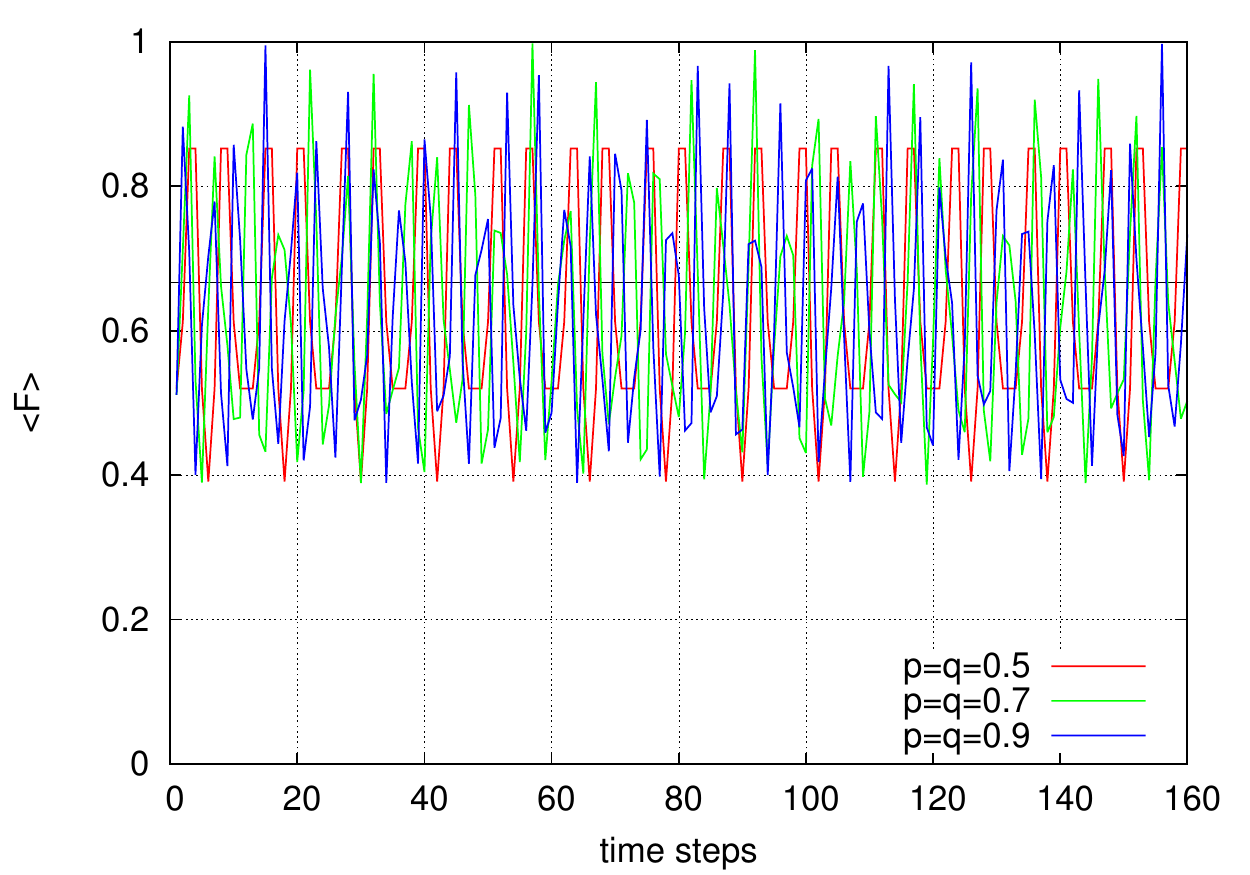}}
\subfigure[]{\includegraphics[scale=0.66]{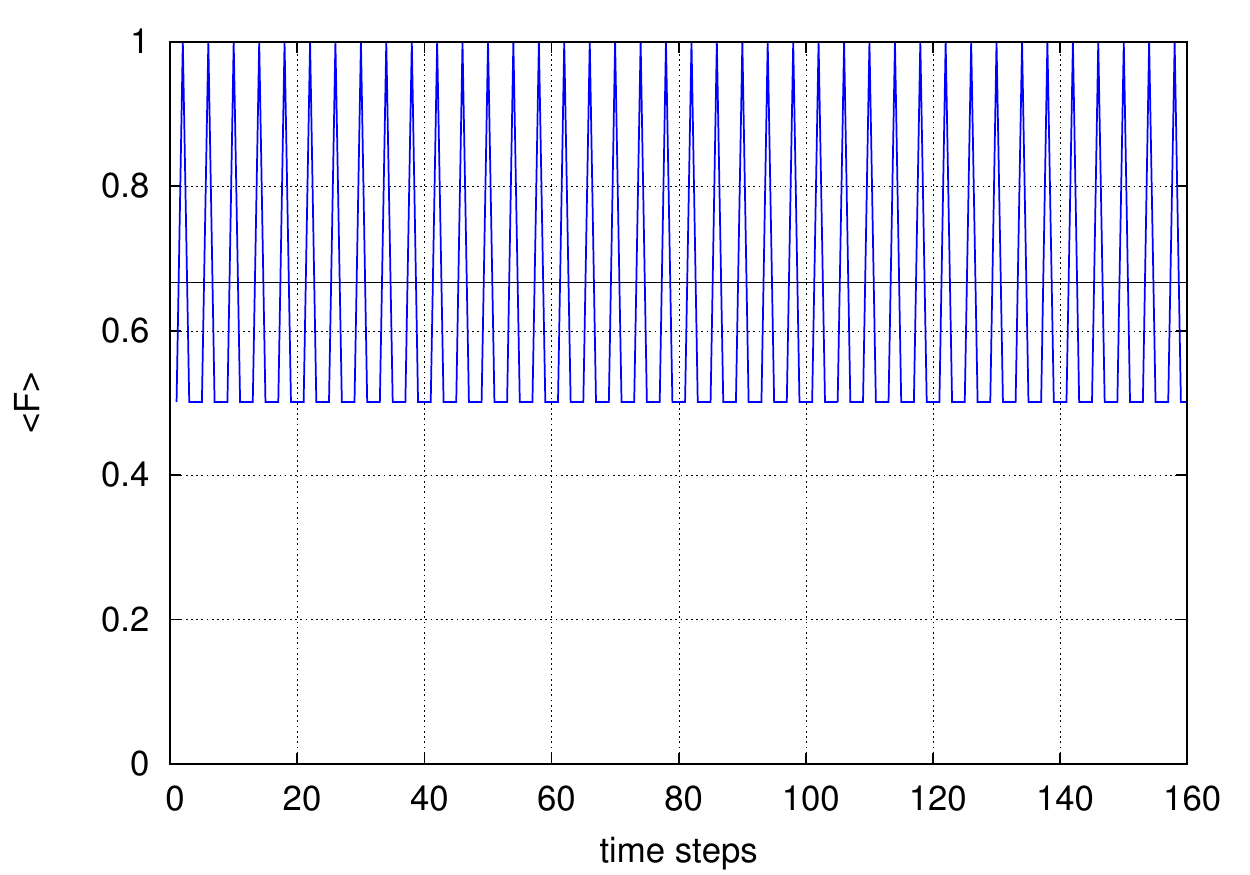}}
\caption{Average fidelity in time for a chain of $N=8$ qubits, with $\xi=0$,$\{\gr{w}_{\mu}\}=0$,
for (a) a linear chain for different values of $p$ and $q$; (c)
a ring for $p=q=(0.5,0.7,0.9)$;
the two extreme cases of a dynamics driven by an almost maximum amplitude damping
strenght parameter ($p=1$, $q=\epsilon$; in the plots $\epsilon=5\cdot 10^{-3}$) and a swap channel
($p=q=1$) for a linear chain (b) and a ring (d) (where the two resulting trends in $\langle F(t)\rangle$
are perfectly superimposed).}
\label{fig:pqChain}
\end{figure*}

An even more dramatic effect on the transport performance is evident when comparing 
settings in which the main two dynamical parameters $p$ and $q$ take equal or different
values. As Fig.\ref{fig:pqChain} shows, when $p=q$ the wide oscillations 
allow for high fidelity peaks, whereas for $p\neq q$ these oscillations are quickly damped
to a steady state close to $1/2$. This is ascribable to the fact that turning on the amplitude
damping component ($\eta\neq 0$) of the channel means that coherences in the SES get suppressed by
a factor $\sqrt{1-\eta}$ at each time step and that should affect the transport of initial coherences with
the vacuum state as well. 
%%%%%%%%%%%%%%%%%%%%%%%%%%%%%%%%%%% FIG
%%%%%%%%%%%%%%%%%%%%%%%%%%%%%%%%%%%
In the limit of maximum amplitude damping ($p=1,q=\epsilon$, $\epsilon\simeq 0$),
the channel does not allow for coherences to build up in time in the SES. However, here comes the science bit: 
in such a limit the transfer is almost perfect and the excitation is locked at the end of the chain, 
as it can be easily understood by recalling that for vanishingly small values of $q$ 
the excitation cannot travel back along the chain. The steady classical maximum
value of $\langle F(t)\rangle$ in Fig.\ref{fig:pqChain}(b) is due to the fact that initial vacuum coherences
are perfectly transported at the beginning till the end of the chain and then progressively 
absorbed, as one can realize 
by directly computing the action of the channel in this particular case.

At the opposite extremum, when $p=q=1$ (and $\xi=0$), the channel is reduced to the sole action of 
$K_0^{(l)}$, which for $\phi_1=\phi_2=0$ is a swap channel. The peak structure that can be
observed in Fig. \ref{fig:pqChain}(b) is obtained because the swap channel makes the entries $\tilde{\rho}_{11}$
bounce back and forth along the diagonal and similarly moves back and forth the two coherences
along their row(column), as we can see by directly computing the first time step evolution
on a three qubit QCA (with $\gr{W}_{\mu}=0, \; \forall \mu$):
\begin{small}
\be
\label{eq:swap}
\tilde{\rho}(0)=
\begin{bmatrix}
\rho_{00} & \rho_{10} & 0 & 0 \\ 
\rho_{01} & \rho_{11} & 0 & 0 \\
0 & 0 & 0 & 0 \\
0 & 0 & 0 & 0 \\
\end{bmatrix} \Rightarrow 
\tilde{\rho}(1)=
\begin{bmatrix}
\rho_{00} & 0 & 0 & |c_0|^2\rho_{10} \\ 
0 & 0 & 0 & 0 \\
0 & 0 & 0 & 0 \\
|c_0|^2\rho_{01} & 0 & 0 & \rho_{11} \\
\end{bmatrix},
\ee  
\end{small}
The result above is obtained without specifying a priori any values for the tuple $\{c_{\mu}\}$.
As already anticipated at the beginning of this section, when the dynamics is fully unitary
($K^{(l)}_1=K_2^{(l)}=0$) only terms proportional to $c_0$ will survive in the first row(column)
of the global density matrix $\tilde{\rho}(t)$, hence choosing $\{c_{\mu}\}=(1,0,0)$
is crucial in order to have perfect state transfer. \\

Let us now add periodic boundary condition to the chain. Numerics show that two
considerations can be made about the differences in quantum state transfer performance
between this ring-like configuration and the previous one.
The first one is specific to the case of maximum amplitude damping $\eta\simeq 1$
where, as opposed to what happened in an open chain, now we find that vacuum coherences
are not absorbed after a transient time but are perfectly periodically transported 
back and forth along the chain together with the excitation, giving rise to exactly
the same maximum peaks structure resulting from a \textquotedblleft swap\textquotedblright
dynamics (Fig.\ref{fig:pqChain}(d)). Secondly, we note that in general, when equal settings are
compared, introducing periodic boundary conditions considerably improoves the state
transfer's fidelity. As an example of this general trend, compare 
Fig.\ref{fig:pqChain}(c) and \ref{fig:pqChain}(a).     

%%%%%%%%%%%%%%%%%%%%%%%%%%%%%%%%%%%%%%%%%%%%%%%%%%%%%%%%%%%%%%%%%%%%%%%% NOISE
\subsection{The noise}
\begin{figure}[htb!]
\includegraphics[scale=0.66]{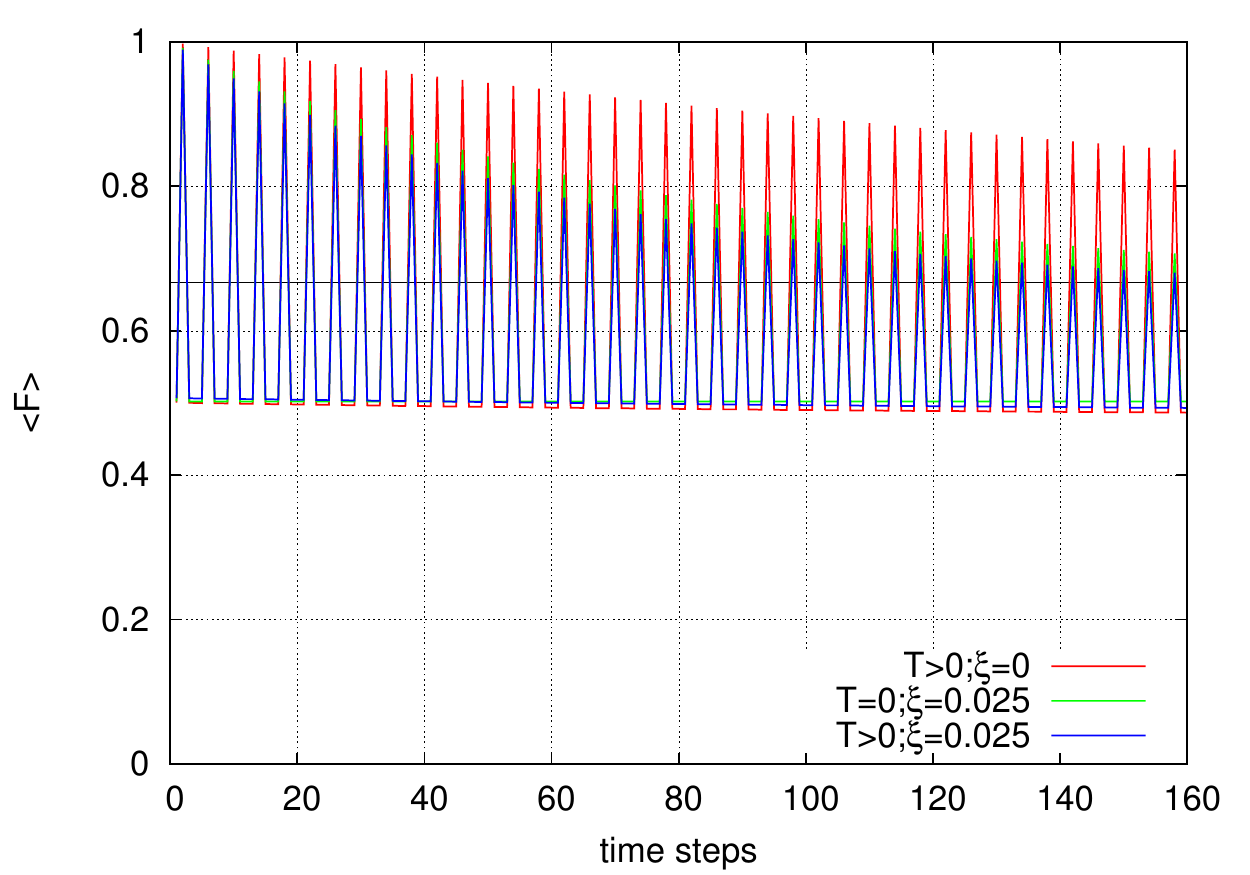}
\caption{Noisy dynamics for a ring of $N=8$ qubits and $p=q=1$. 
In the depicted case $T>0$ corresponds to the value $T=0.1$
(see Eq.(\ref{eq:T})).
However, the magnitude chosen for the two parameters $\xi$ and $T$ does not acquire any particular meaning here,
as the figure is intended to be only a representative example of the damping in $\langle F(t)\rangle$
resulting from introducing noise  (cfr. Fig.\ref{fig:pqChain}(d)).}
\label{fig:noise}
\end{figure}

Finally, let us conclude this section by discussing the role of noise in our system. 
In our dynamics the system-environment coupling is modeled through the action of three 
different general kinds of noise: amplitude damping, phase damping and \apicisx excitation pumping\apicidx
-like noise.  
When transferring excitations only (SES), the former two could represent an advantage over 
a pure unitary dynamics. In fact, the amplitude damping is needed in order to introduce an
asymmetry between the two direction of propagation, acting as a driving force to draw
the excitation towards the end of the chain, whereas the introduction of some dephasing
may smooth the localization effect previously discussed (Fig.\ref{fig:localization}(a),(b)),
allowing for a better transfer \cite{papero1}.
For quantum state transfer though, coherences play a fundamental role and, as expected,
all the sources of noise we considered are always (apart from the very specific case of Fig.\ref{fig:pqChain}(d)) detrimental.
More specifically, amplitude damping and dephasing result in a damping
of the coherent oscillations of $\langle F(t)\rangle$ (see Fig. \ref{fig:pqChain}(a) and \ref{fig:noise}),
whereas the restrictions imposed by the causality condition on $\{\gr{w}_{\mu}\}$
and $\{z_{\mu}\}$ (last two lines of Eqs.(\ref{eq:allconstraints}))
make a quantum state transfer dynamics with $\{\gr{w}_{\mu}\}\neq 0$ strongly unfavourable. 
Assuming either $\{z_{\mu}\}\in \mathbb{R}$ or $\{\gr{w}_{\mu}\}\in \mathbb{R}$
(the case where both belong to $\mathbb{C}$ only complicates the picture without changing the results),
we have that the only solutions to Eq.(\ref{eq:allconstraints}) where $\{\gr{w}_{\mu}\}\neq 0$
are such that $z_{\mu}=z$, $\forall \mu$, and we find that this condition, whatever the other parameters
of the dynamics may be, always translates into very poor quantum state transfer performance. \\

For the sake of the argument, one can think of relaxing the causality requirement.
In that case, as the first line in Eqs.(\ref{eq:allconstraints}) is just a matter of normalization,
one can build the vectors by satisfying $\sum_{\mu}{K_{\mu}^{(l)}}^{\dagger}\cdot \gr{w}_{\mu}=0$, 
which, upon substituting Eqs.(\ref{eq:k0},\ref{eq:k1},\ref{eq:k2}), reads:  
\be
\left\{
\begin{array}{lr}
%\sqrt{1-\xi/2}\sqrt{1-\eta}w_0^0+\sqrt{\xi/2}\sqrt{1-\eta}w_1^0+\sqrt{\eta}w_2^1=0 & \\
%\sqrt{1-\xi/2}w_0^1-\sqrt{\xi/2}w_1^1=0 & \\
\sqrt{1-\frac{\xi}{2}}\sqrt{1-\eta}w_0^0+\sqrt{\frac{\xi}{2}}\sqrt{1-\eta}w_1^0+\sqrt{\eta}w_2^1=0 & \\
\sqrt{1-\frac{\xi}{2}}w_0^1-\sqrt{\frac{\xi}{2}}w_1^1=0 & \\
\end{array}
\right. 
\label{eq:wconstr}.
\ee
where we introduced a notation such that $w^a_{\mu}$ stands for the component
$a$, $a=(0,1)$ of each local vector pertaining to the Kraus operator $\mu$. 
Of course, a whole set of solutions for Eqs.(\ref{eq:wconstr}) is possible.
In order not to give too big a bias among the different $\tilde{K}_{\mu}$
driving the dynamics,
it is reasonable to privilege the most \textquotedblleft balanced\textquotedblright
solutions, the ones where $|w_{\mu}^a-w_{\nu}^b|\leq d, \forall \mu,\nu,a,b$, with the smallest possible $d$.
To this aim, it is thus necessary to avoid solutions proportional to 
$(\xi/2)^{-1/2}$, $(1-\eta)^{-1/2}$ or $\eta^{-1/2}$,
where for some regimes of the dynamics some components $w_{\mu}^a$ could be arbitrarily big.
One such solution reads:
\be
\label{eq:T}
\left\{
\begin{array}{lr}
w_0^0=w_0^1=-\frac{T}{\sqrt{3N}}\sqrt{\frac{\xi/2}{1-\xi/2}} & \\
w_1^0=\frac{T}{\sqrt{3N}}(1-\sqrt{\eta}); \quad w_1^1=-\frac{T}{\sqrt{3N}} & \\
w_2^0=w_2^1=\frac{T}{\sqrt{3N}}\sqrt{\xi/2}\sqrt{1-\eta}
\end{array}
\right.
\ee
Above, we assumed $\{\gr{w}_{\mu}\}\in \mathbb{R}$ and we arbitrarily chose a normalization proportional
to $T/\sqrt{N}$,
where the free parameter $T$ tunes the strength of the noise.
Notice that Eqs.(\ref{eq:T}) assure that $\{w_{\mu}^a\}\in \left[\pm T/\sqrt{3N}\right]$.
As a result of relaxing causality, the components of $\{\gr{w}_{\mu}\}$ and $\{z_{\mu}\}$
are now \apicisx decoupled\apicidx and a tuple $\{z_{\mu}\}=(c,0,0)$ - which allows for
the best quantum state transfer performance - can be selected. 
However, as it can be appreciated in Fig.\ref{fig:noise}, the effect of this kind of noise
is very similar to that of the dephasing.
%%%%%%%%%%%%%%%%%%%%%%%%%%%%%%%%%%%%%%%
%%%%%%%%%%%%%%%%%%%%%%%%%%%%%%%%%%%%%%%

%\clearpage
%%%%%%%%%%%%%%%%%%%%%%%%%%%%%%%%%%%%%%%%%%%%%%%%%%%%%%%%%%%%%%%%%%%%%%%%%%%%%%%%% CONCLUSION
\section{Conclusion} \label{sec:conclusion}
We have defined a quantum cellular automaton dynamics to model quantum state transfer
on a chain of qubits. We began by limiting the dynamics to the first excitation sector,
where we established a framework in which classical stochastic processes are embedded
within the formalism of quantum operations. We then enlarged the Hilbert space to
allow for transport of initial coherences as well, finding 
the conditions selecting the class of proper - causal and translational invariant - 
noisy QCA out of all possible dynamics in the $0\oplus 1$ sector. 
The extension of the automata to the vacuum state allows for the possibility to consider the effect of
the quantum to classical transition on on-site coherences. 
As shown through numerical simulations, there are instances of the dynamics in which a transfer fidelity larger than the classical threshold is achievable.
We have then discussed the role played by the different parameters characterizing the dynamics. In particular we 
show the existence of an optimal choice of the free phases $(\phi_1,\phi_2)$ parametrizing the local evolution, that allows for perfect state transfer. We then highlighted the different impact of the sources of noise taken into account, in particular amplitude damping and dephasing, on excitation transfer and quantum state transfer: while in the first case, as described in \cite{papero1}, noise can sometimes be beneficial, for the latter it proves always detrimental.
Note that the capability of transferring a quantum state with high fidelity mirrors the 
ability to transmit quantum entanglement, so that the possibility to distribute entanglement 
between different lattice sites is implied in our study. 
From a more general perspective, our results suggest that QCA are
a new and potentially interesting alternative scheme to model relevant processes for
quantum communication.

%%%%%%%%%%%%%%%%%%%%%%%%%%%%%%%%%%%%%%%%%%%%%%%%%%%%%%%%%%%%%%%%%%%%%%%%%%%%%%%%% ACKNOWLEDGEMENTS
\section*{Acknowledgments}
MA and AS acknowledge joint financial support from Mischa Stocklin and UCL through the 
Impact scholarships scheme; AS and MGG acknowledge support from EPSRC through grant EP/K026267/1. 
Many thanks go to the anonymous reviewer that prompted us to examine 
the causality conditions on the extended CP-maps.
We also thank Martin Plenio and Mischa Stocklin for useful discussion, 
as well as Tedeschi for his unfailing support.

%%%%%%%%%%%%%%%%%%%%%%%%%%%%%%%%%%%%%%%%%%%%%%%%%%%%%%%%%%%%%%%%%%%%%%%%%%%%%%%%% BIBLIOGRAPHY
\bibliographystyle{prsty}

%%%%%%%%%%%%%%%%%%%%%%%%%%%%%%%%%%%%%%%%%%%%%%%%%%%%%%%%%%%%%%%%%%%%%%%%%%%%%%%%%

\appendix %%%%%%%%%%%%%%%%%%%%%%%%%%%%%%%%%%%%%%%%%%%%%%%%%%%%%%%%%%%%%%%%%%%%%%% APPENDICES

\section{Partial trace in the $0\oplus 1$ sector.} \label{appendixA} %%% PARTIAL TRACE
Here we show how to calculate a partial trace in the $\mathcal{H}_0\oplus \mathcal{H}_1$ sector of the
Hilbert space, assuming that no other sectors are populated.
The Hilbert space is spanned by the basis $\{\ket{n},0\le n\le N\}$,
where $\ket{0}$ is the global vacuum and $\ket{n>0}$ represents the state with the $n^{\rm th}$ qubit
in the excited state and all the other qubits in the ground state.
In order to obtain the reduced state of qubit $x$, a generic state of the global system 
$\tilde{\rho}=\sum_{j,k=0}^N \psi_j \psi_k^* \ketbra{j}{k}$ can be rewritten in a convenient way
bearing in mind that whenever the excitation is at site $x$ it cannot be anywhere else ($\mathcal{L}/x$), and vice versa:
%\begin{small}
\bea
\sum_{j,k=0}^N \psi_j \psi_k^* \ketbra{j}{k} & = & \sum_{\begin{array}{c} \scriptstyle j,k=1 \\ \scriptstyle (j,k\neq x)\end{array}} \psi_j\psi_k^*\ketbra{j}{k}_{\mathcal{L}/x}\otimes \ketbra{0}{0}_x + \psi_0\psi_k^*\ketbra{0}{k}_{\mathcal{L}/x}\otimes \ketbra{1}{0}_x + \psi_j\psi_0^*\ketbra{j}{0}_{\mathcal{L}/x}\otimes \ketbra{0}{1}_x \\ \nonumber
 & + & |\psi_x|^2\ketbra{0}{0}_{\mathcal{L}/x}\otimes \ketbra{1}{1}_x +\psi_x\psi_0^*\ketbra{0}{0}_{\mathcal{L}/x}\otimes \ketbra{1}{0}_x + \psi_0\psi_x^*\ketbra{0}{0}_{\mathcal{L}/x}\otimes \ketbra{0}{1}_x \; . 
\eea
%\end{small}
It is now easy to see that a partial trace over the lattice $\mathcal{L}/x$ gives:
\be
\rho_x=\sum_{l\neq x}|\psi_l|^2\ketbra{0}{0}+|\psi_x|^2\ketbra{1}{1}+\psi_x\psi_0^*\ketbra{1}{0}+\psi_0\psi_x^*\ketbra{0}{1} \; ,
\ee
which is a well defined, trace-one reduced state.

\section{QCA translational invariance in the $0\oplus 1$ sector.} \label{appendixB} %%% TRANSL. INVARIANCE
The $2$-site block translational invariance condition can be checked by directly computing the 
sum of the commutators between the squared lattice shift operator $\sigma^2$ and the operators defining the QCA map:  
\begin{small}
\begin{align}
%\be
\sum_{\mu}\left[\tilde{\sigma^2},\tilde{K_{\mu}}\right] = \sum_{\mu}\left\{
\begin{bmatrix}
1 & \gr{0} \\
\gr{0} & \sigma^2 \\
\end{bmatrix}
\begin{bmatrix}
z_{\mu} & \gr{0} \\
\gr{W}_{\mu} & K_{\mu}^{(s)} \\ 
\end{bmatrix} \; - \;
\begin{bmatrix}
z_{\mu} & \gr{0} \\
\gr{W}_{\mu} & K_{\mu}^{(s)} \\
\end{bmatrix}
\begin{bmatrix}
1 & \gr{0} \\
\gr{0} & \sigma^2 \\
\end{bmatrix}
\right\} =
\begin{bmatrix}
0 & \gr{0} \\
\sum_{\mu}\left( \sigma^2\cdot \gr{W}_{\mu}-\gr{W}_{\mu}\right) & \sum_{\mu}\left[ \sigma^2,K_{\mu}^{(s)}\right] \\
\end{bmatrix} \; .
%\ee
\end{align}
\end{small}
In the SES with periodic boundary conditions, the block diagonal structure of $K_{\mu}^{(s)}$ ensures that 
$\sum_{\mu}\left[ \sigma^2,K_{\mu}^{(s)}\right]=0$. The translational invariance of the extended QCA
dynamics is thus retained when the $\gr{W}_{\mu}$ vectors are translationally invariant, as in that case
the order with which a squared shift of the lattice and the QCA map are applied does not matter:
\be
\sum_{\mu \nu}\tilde{\sigma}^2\left( \tilde{\sigma}\tilde{K_{\mu}}\tilde{\sigma}^{\dagger}\tilde{K_{\nu}}\tilde{\rho}\tilde{K_{\nu}}^{\dagger}\tilde{\sigma}\tilde{K_{\mu}}^{\dagger} \tilde{\sigma}^{\dagger}\right) \tilde{\sigma}^{2\dagger}=\sum_{\mu \nu} \tilde{\sigma}\tilde{K_{\mu}}\tilde{\sigma}^{\dagger}\tilde{K_{\nu}}\left(\tilde{\sigma}^2\tilde{\rho}\tilde{\sigma}^{2\dagger}\right)\tilde{K_{\nu}}^{\dagger}\tilde{\sigma}\tilde{K_{\mu}}^{\dagger} \tilde{\sigma}^{\dagger} \tilde{\sigma}^{2\dagger} \; .
\ee  
 
\section{QCA causality in the $0\oplus 1$ sector.} \label{appendixC} %%% CAUSALITY
Given a global density matrix $\tilde{\rho}=\sum_{j,k=0}^N \rho_{jk} \ketbra{j}{k}$, we have (See Appendix \ref{appendixA})
that the reduced state of a qubit at any site $x$ and the reduced state of the neighborhood $\mathcal{N}_x$ ($\mathcal{N}_x=\{x, \; y\equiv x+1\}$)
are, respectively:
\begin{align} \label{eq:reduced}
\rho_x =
\begin{bmatrix}
\sum_{l\neq x}\rho_{ll} & \rho_{0x} \\
\rho_{0x}^* & \rho_{xx}
\end{bmatrix} \; , \quad
\rho_{\mathcal{N}_x} =
\begin{bmatrix}
\sum_{l\neq (x,y)}\rho_{ll} & \rho_{0x} & \rho_{0y} \\
\rho_{0x}^* & \rho_{xx} & \rho_{xy} \\
\rho_{0y}^* & \rho_{xy^*} & \rho_{yy} \\
\end{bmatrix} \; .
\end{align}
In order to check the causality condition Eq.(\ref{eq:causality}), it is sufficient to directly calculate the reduced state
of qubit $x$ after an application of half step of the QCA, the step where the lattice partition is such that the
the two qubits in $\mathcal{N}_x$ interact through one of the local maps composing the global dynamics of Eq.(\ref{eq:QCAmap}).
After a straightforward calculation, we find that the new state of qubit $x$ can be expressed in the form:
\begin{align} 
\rho_{x}^{(new)}= 
\begin{bmatrix}
1-f_{x|2}(\rho_{0x},\rho_{0y},\rho_{xx},\rho_{yy},\rho_{xy})-\rho_{00}\sum_{\mu}w_{\mu}^{x|2} & g_{x}(\rho_{0x},\rho_{0y}) + \rho_{00}\sum_{\mu}z_{\mu}w_{\mu}^{x|2} \\
c.c & f_{x|2}(\rho_{0x},\rho_{0y},\rho_{xx},\rho_{yy},\rho_{xy}) + \rho_{00}\sum_{\mu}w_{\mu}^{x|2}\\
\end{bmatrix} \; ,
\end{align}
where $\bullet |2$ means $\bullet$ modulo $2$.
The value of functions $f$ and $g$ is determined only by the parameters of the dynamics 
(through $\{K_{\mu}\},\; \{\gr{W}_{\mu}\}\; \mbox{and }\{z_{\mu}\}$) and the components of the reduced state of
the neighborhood before the evolution (Eq.(\ref{eq:reduced})). 
Two global states $\tilde{\rho}$ and $\tilde{\rho}'$ having $\rho_{\mathcal{N}_x}=\rho_{\mathcal{N}_x}'$ before
the evolution will thus give new reduced states on $x$ such that:
\begin{align} \label{eq:pippo}
\rho_x^{(new)}-\rho_x^{'(new)} =
\begin{bmatrix}
(\rho_{00}'-\rho_{00})\sum_{\mu}w_{\mu}^{x|2} & (\rho_{00}-\rho_{00}')\sum_{\mu}z_{\mu}w_{\mu}^{x|2} \\
c.c & (\rho_{00}-\rho_{00}')\sum_{\mu}w_{\mu}^{x|2} \\
\end{bmatrix}
\end{align}
In order to have causality, the above difference must be zero for all $\rho_{00}$ and $\rho_{00}'$. 

We thus find that, when preserving the dynamics Eq.(\ref{eq:QCAmap}) in the single excitation sector, the class of causal
QCA is selected by the following constraints:
\be
\label{eq:newconstraints}
\left\{
\begin{array}{lr}
\sum_{\mu}z_{\mu}w_{\mu}^0=\sum_{\mu}z_{\mu}w_{\mu}^1=0 & \\
\sum_{\mu}w_{\mu}^0=\sum_{\mu}w_{\mu}^1=0 & \\
\end{array}
\right.
\ee

\end{document}